\documentclass{mn2e}
\usepackage{graphicx}

\title[Structure and conditions of L1544]
{Constraining the structure of the non-spherical preprotostellar
core L1544}
\author[S. D. Doty et al.]
       {Steven D. Doty$^1$, Sheila E. Everett$^1$,
	Yancy L. Shirley$^2$\thanks{Jansky Postdoctoral Fellow},
	\newauthor
	Neal J. Evans II$^3$, 
	and Matthew L. Palotti$^{1,4}$ \\
        $^1$Department of Physics and Astronomy, Denison University,
	Granville, OH  43023, USA\\
	$^2$National Radio Astronomical Observatory, Socorro, NM,
	87801, USA\\
	$^3$Department of Astronomy, University of Texas at Austin,
	Austin, TX  78712-1083, USA\\
	$^4$Department of Physics, University of Wisconsin-Madison,
	Madison, WI  53706-1390, USA
	} 
\date{Accepted 31 January, 2005. 
      Received ; 
      in original form  }

\pagerange{\pageref{firstpage}--\pageref{lastpage}}
\pubyear{2004}

\begin{document}

\maketitle

\label{firstpage}

\begin{abstract}
We present a study of the pre-protostellar core L1544.  
A series of self-consistent, three-dimensional continuum 
radiative transfer models are constructed.  The outputs of these
models are convolved with appropriate telescope beam 
responses, and compared with existing SCUBA data.
The resulting comparison allows us to constrain the structure
of L1544.  We find that the source is well-fit by a
prolate spheroid, having an ellipsoidal power-law density
distribution of index $m \sim 2$ ($1.75 < m < 2.25$) in to
at least $r \sim 1600$ AU.  For $r<1600$ AU, the data are 
consistent with either an extension of the power law to smaller
radii, or a flattened (Bonner-Ebert like) density distribtion.
Furthermore,
we find an optical depth along the short axis at $1300 \mu$m of
$\tau_{1300,\mathrm{short}}=5\times10^{-3}$  
($2 \times 10^{-3} < \tau_{1300,\mathrm{short}} < 8 \times 10^{-3}$),
a central luminosity $L_{*}=0$ ($< 10^{-3}$ L$_{\odot}$), a 
long axis diameter
$D=0.1$ pc ($0.08 < D(\mathrm{pc}) < 0.16$ ;   
$16000 < D(\mathrm{AU}) < 32000$), an axis ratio
$q=2$ ($1.7 < q < 2.5$), and
an external ISRF defined by Mathis, Mezger, \& Panagia (1983) to
within 50 per cent.
The outer diameter and axis ratio may each be somewhat larger due
to potential on-source chopping in the observations, 
and the projection of the long axis onto the plane of the sky.  
While these results are similar to those
inferred directly from observations or spherical modeling due to the
source transparency at submillimeter wavelengths, 
we infer a smaller size, lower mass, and higher
optical depth / column density, exposed to a stronger external
radiation field than previously assumed.  Finally, we find that both the 
spectral energy distribution (SED) 
and surface brightness distribution are necessary to 
constrain the source properties in this way, and even a modest
variation in $\chi^{2}$ can significantly alter the fit quality.
\end{abstract}


\section{Introduction}

The structure of bound, starless cores (e.g., 
Myers, Linke, \& Benson 1983; 
Benson \& Myers 1989) is not well known.  This structure is
important, as it should drive the ensuing
evolution of the core, and give insight into the initial
conditions of star formation free from the disruption of
a central protostar.  Ward-Thompson, Scott, Hills, \& Andr\'{e}
(1994) detected millimeter and submillimeter continuum 
radiation from dust in these regions, and dubbed them
preprotostellar cores (PPCs).   

Following this detection, significant work has continued 
toward determining the conditions associated with these regions.
Ward-Thompson, Scott, Hills, \& Andr\'{e} (1994) inferred that
the density distributions do not follow isothermal sphere
distributions ($n(r) \propto r^{-2}$) over a range of cores, 
a result supported by various other studies 
(e.g. Andr\'{e}, Ward-Thompson, \& Motte 1996; 
Ward-Thompson, Motte, \& Andr\'{e}
1999; Bacmann et al. 2000).  These studies utilized semi-analytic
techniques to infer the dust density distribution.  Similar techniques
such as colour temperature variation, were used by 
Ward-Thompson, Andr\'{e}, \& Kirk (2002) (along with ISO flux data)
to infer the temperature
distribution and to place constraints on the luminosity of the
central source.  While many of these semi-analytic approaches
yield physically meaningful information, the
specifics of the approach can affect the reliability of the
conclusions (Doty \& Palotti 2002).  

Submillimeter (submm) data for many PPCs were obtained from the Submillimeter
Common-User Bolometric Array (SCUBA) on the James Clerk Maxwell
Telescope (JCMT) by Shirley et al. 2000.  In an analysis of these data, 
Evans et al. (2001) assumed spherical symmetry and utilized
a self-consistent dust radiative transfer model to infer the
properties and structure of their source sample.  
They found that Bonner-Ebert (BE) spheres provided good fits
to the data, with L1544 requiring the highest central
density, $n_c = 10^6$ cm$^{-3}$, in fact.
However, a power-law density distribution could not be ruled
out when a self-consistent dust temperature variation was included.
This result, while intriguing, 
relies upon the assumed underlying source geometry.
Unfortunately, the observed isophotes for L1544 are decidedly 
non-spherical, leading to potential uncertainty in these conclusions.

In this paper we report on a study comparing existing data
with detailed, three-dimensional, continuum radiative transfer models 
to constrain the conditions associated with L1544.  In section two,
we describe L1544 and previously inferred properties. 
We briefly describe the model and range of parameter space considered
in section three.  In section four, we compare the models with observations
in constraining the source properties.  Finally, we conclude in section five.

\section{L1544
              }

L1544 is located in Taurus at a distance of 140 pc (Elias 1978).
The source has been well-observed.  There is no reported
IRAS source associated with the core (Ward-Thompson, Scott, Hills, 
\& Andr\'{e} 1994).  However, submillimeter
data have been obtained at the JCMT 
at 450, 850, and 1100 $\mu$m with a single element
by Ward-Thompson, Scott, Hills, \& Andr\'{e} (1994), and
using SCUBA (Shirley et al. 2000).  Further submillimeter data
were taken at 1.3-mm using IRAM 
(Ward-Thompson, Scott, Hills, \& Andr\'{e} 1994; 
Ward-Thompson, Motte, \& Andr\'{e} 1999).   
L1544 has also been studied with the
\textit{Infrared Space Observatory}.  In particular, Bacmann
et al. (2000) utilized ISOCAM imaging data ($\sim 7 \mu$m),
while Ward-Thompson, Andr\'{e}, \& Kirk (2002) utilized the ISOPHOT
photometer at 90, 170, \& 200 $\mu$m.

In Table 1, we reproduce the observed fluxes that we adopt in 
our study.  We choose the 450, 850, and 1300 $\mu$m data from 
Shirley et al. (2000), as the results are from an array, rather
than a single element bolometer.  
The 90, 170, \& 200 $\mu$m fluxes are taken 
from Ward-Thompson, Andr\'{e}, \& Kirk (2002).  In both cases,
the uncertainties in Table 1 are the quoted statistical 
uncertainties added in quadrature with an assumed 30 per cent
calibration uncertainty in the data.

   \begin{table}
      \caption[]{Fluxes toward L1544}
         \label{l1544fluxes}
         \begin{tabular}{rrrr}
            \hline
            Wavelength & Flux Density & Uncertainty$^a$ & ref \\ 
	    ($\mu$m) & (Jy) & { } & { } \\
            \hline
	     90 & $<0.57^{b}$& n/a & 1 \\
	    170 & 14.9$^{b}$ & $(0.2,4.5) = 4.5$ & 1 \\
	    200 & 18.6$^{b}$ & $(0.3,5.6) = 5.6$ & 1 \\
	    450 & 17.4$^{c}$ & $(6.7,3.5) = 7.5$ & 2 \\
	    850 & 3.64$^{c}$ & $(0.18,0.73) = 0.75$ & 2 \\
	   1300 & 0.27$^{d}$ & $(0.04,0.05) = 0.07$ & 2 \\
	   
            \hline
     \end{tabular}\\
\begin{list}{}{}{}
\item[$^{\mathrm{a}}$] (Statistical, 30\% cal.) 
                             uncertainties added in quadrature  
\item[$^{\mathrm{b}}$] Flux density in a 150 arcsec aperature  
\item[$^{\mathrm{c}}$] Flux density in a 120 arcsec aperature   
\item[$^{\mathrm{d}}$] Flux density in a 40 arcsec aperature 
\item[$^{\mathrm{1}}$] Ward-Thompson, Andr\'{e}, \& Kirk (2002)  
\item[$^{\mathrm{2}}$] Shirley, Evans, Rawlings, \& Gregersen (2000) 
\end{list}
   \end{table}

The spectral energy distribution (SED) can help constrain the
density distribution.  However, the constraint is not unique
(Butner et al. 1991; Men'shchikov \& Henning 1997; Doty \& Palotti
2002).  On the other hand, the spatial distribution of intensity on the
sky in the form of maps provides significantly more data with which to
constrain the models (Adams 1991; Ladd et al. 1991).  As a result, we
also consider the 450 $\mu$m and 850 $\mu$m intensity maps 
produced by Shirley, et al. (2000).  



The data above were analyzed in detail in the presenting papers 
in order to constrain the structure and conditions associated with
L1544.  While these approaches were semi-analytic in 
nature and/or assumed spherical symmetry, they represent the norm
in analyzing long wavelength emission from dust.  Consequently, 
these results provide a useful first-order approximation to the 
underlying source conditions, and are presented in Table 2.

   \begin{table}
      \caption[]{Previously inferred conditions for L1544}
         \label{l1544conditions}
         \begin{tabular}{lrr}
            \hline
            Parameter & Value & ref \\ 
            \hline
	     Optical depth at $1300 \mu$m ($\tau_{1300}$) & $2\times10^{-3}$ & 3\\
	     Optical depth at $200 \mu$m ($\tau_{200}$) & 0.06 & 5 \\
	     Density distribution & BE$^{a}$ or PL($m=2$)$^{b}$ & 2 \\
	     ISRF ($G_{\mathrm{MMP}}^{c}$) & $0.6$ & 6 \\
             Bolometric Luminosity ($L_{\mathrm{bol}}$) [$L_{\odot}$] & 1.0 $\pm$ 0.3 & 1 \\
             Central Luminosity ($L_{\mathrm{central}}$) [$L_{\odot}$] & $< 0.1$ & 5 \\
             Diameter ($D$) [$\times 1000$ AU] & $\sim 17.8$ & 4 \\	     
             Central mass$^{d}$ ($M_{120}$) [$M_{\odot}$] & $0.4^{+0.8}_{-0.3}$ & 1$$ \\
	     Total mass ($M_{\mathrm{tot}}$) [$M_{\odot}$] & 2.7 $\pm$ 0.7 & 2 \\ 
             H$_{2}$ column ($N_{\mathrm{H}_{2}}$) [$10^{22}$ cm$^{-2}$] & $ 6-13$ & 3,5 \\
            \hline
     \end{tabular}\\
\begin{list}{}{}{}
\item[$^{\mathrm{a}}$] Bonner-Ebert sphere 
\item[$^{\mathrm{b}}$] Power-law of the form $n(r)=n_{0}(r_{0}/r)^{m}$  
\item[$^{\mathrm{c}}$] ISRF strength relative to standard MMP  
\item[$^{\mathrm{d}}$] In a 120 arcsec diameter beam  
\item[$^{\mathrm{1}}$] Shirley, Evans, Rawlings, \& Gregersen (2000)  
\item[$^{\mathrm{2}}$] Evans, Rawlings, Shirley, \& Mundy (2001)  
\item[$^{\mathrm{3}}$] Ward-Thompson, Motte, \& Andr\'{e} (1999)  
\item[$^{\mathrm{4}}$] Bacmann et al. (2000)  
\item[$^{\mathrm{5}}$] Ward-Thompson, Andr\'{e}, \& Kirk (2002)  
\item[$^{\mathrm{6}}$] Young, Shirley, Evans, \& Rawlings (2003)
\end{list}
   \end{table}

\section{Model}
We have constructed detailed, self-consistent, three-dimensional
radiative transfer models through dust.   The model utilizes
a monte-carlo approach combined with an approximate lambda
iteration to ensure true convergence even at high optical
depths.  The model has been tested against existing 1-D 
(Egan, Leung, \& Spagna, 1988); and 2-D (Spagna, 
Leung, \& Egan 1991) codes with good success.
 
Based upon the
input parameters discussed below, we solve for the dust
temperature and radiation field at each point in the
model cloud with a typical resolution of $\sim 10^{15}$ cm.  
The emergent radiation, after subtraction of the 
background ISRF, is then convolved with the
appropriate telescope beam (Shirley et al. 2000)
for comparison with observations.
We do not simulate the chopping, as the chop direction and
size varied with time during the observations.
Evans et al. (2001) found that chopping affected the radial
profile in the outer regions and were unable to determine 
an outer radius.  While we will give an outer radius in this
paper, it should be interpreted with caution.  We have simulated
the contours only out to radii that are not strongly affected by
chopping, so conclusions about the shape of the density profile
to that point should be safe.
The approach of simulating the observations 
has the advantage that we can directly compare 
models to observational output, while minimizing the need for
imposing outside approximations/assumptions to the observational
data.

\subsection{Input data}
We adopt a triaxial ellipsoidal density structure for our cloud models, 
of the form
\begin{equation}
n(x,y,z)=n_{0} 
\left[
\sqrt{
\frac{x^{2}}{a^{2}} + \frac{y^{2}}{b^{2}} +
\frac{z^{2}}{c^{2}}
}
\right]
^{-m}
,
\end{equation}
where $n$ is the number density, 
$n_0$ is the reference density, $(x,y,z)$ defines the position,
$(a,b,c)$ are parameters specifying the shape of the ellipsoid,
and $m$ is the dust density distribution exponent discussed in the
following subsection.
This form is consistent with fact that the observed distribution of
projected axial ratios is well matched by randomly oriented,
intrinsically prolate cores (Myers et al. 1991; Ryden 1996).  It is
also consistent with the nonaxisymmetric evolution of magnetically
subcritical cores (Nakamura \& Li 2002).  We follow this work
and the symmetry in the NW-SE direction, and reduce our density
structure to a prolate spheroid having the long ($x$) axis in the
NW-SE direction, and being symmetric in the $y-$ and $z-$ directions.
In this case, we take $b=c$, and define the axis ratio to be
$q \equiv a/b$.  Based upon the observed isophotes, we consider axis
ratios in the range $1.2 < q < 5$.

We assume that the dust density follows a power-law, with 
exponent $m$.  This significantly simplifies the source
parameterization while maintaining the ability to consider
various amounts of central condensation.  
The power-law parameterization is commensurate with similar 
spherical models, as the spherical average density distribution
for any ellipsoid given by equation (1) is 
$n_{\mathrm{spherical}} \propto r^{-m}$, and is  
given by $n_{\mathrm{spherical}}(r) = \int d \Omega
n(x,y,z) / \int d \Omega = n_{0}C(a,b,c,m)r^{-m}$, where
$C(a,b,c,m)$ is a constant that depends only upon the geometry, 
and the density distribution.  
Furthermore, this 
approach is consistent with both observations
(e.g. Evans et al. 2001) and theoretical (e.g. Nakamura \& Li 2002;
Curry \& Stahler 2001; Ciolek \& Basu 2000) models of evolution of
magnetized clouds.  Based upon previous spherical models, we consider
density distribution exponents in the range $1.0 < m < 3.0$.

The outer radius is the distance along the long ($x-$) axis at which 
the cloud structure terminates, and is half of the adopted
long-axis diameter, $D$.  At this point, the cloud is
presumed to mesh with the surrounding, more diffuse, medium.
We consider diameters in the range 
$0.06 < D(\mathrm{pc}) < 0.2$  
($12000 < D(\mathrm{AU}) < 40000$).
The inner radius defined as the
distance from the center of the model space to the edge of the
first non-central cell is $r_{\mathrm{in}} \sim 0.008 \times D$. 

The interstellar radiation field (ISRF) is the primary heat source for the 
grains in most of our models.  We adopt the ISRF compiled
by Mathis, Mezger, \& Panagia (1983; hereafter MMP).  The strength of the ISRF, 
$G_{\mathrm{MMP}} \equiv $ (adopted ISRF) / (MMP ISRF),
is in the range $0.3 < G_{\mathrm{MMP}} < 3.0$.  We also include the potential
effects of a luminous central source by considering 
$10^{-6} < L_{*}(\mathrm{L}_{\odot}) < 10^{-2}$.

The absolute dust density (or mass) is constrained by the 
optical depth to source center along the long axis via the
optical depth at 450 $\mu$m, namely 
$\tau_{450} = \int n(x) \langle Q(450\mu\mathrm{m}) \pi a^{2} \rangle dx$.  
Here $n(x)$ is the number density specified above, $Q$ is the grain 
absorption efficiency at $450$ $\mu$m discussed below,
and $a$ is the grain radius taken to be $0.1$ $\mu$m.  
Since the grains are small compared to the wavelength of the majority of
the incident and re-emitted radiation (having $\lambda > 1 \mu$m), the
grain opacity per unit mass is independent of the grain size 
(Ossenkopf \& Henning, 1994).
Test models with different grain sizes confirm this independence of
our results with adopted grain size.
Finally, we consider
optical depths in the range $0.01 < \tau_{450,\mathrm{long}} < 0.5$.

We adopt the dust opacities in column 5 of Table 1 of 
Ossenkopf \& Henning
(1994).  They have been successful in 
fitting observations of both high-mass (e.g., van der Tak et al. 
1999, 2000; Mueller et al. 2002) 
and low-mass (e.g., Evans et al. 2001) star 
forming regions.  These opacities are calculated for 
grains grown by coagulation and accretion of thin ice mantles
for $10^{5}$ years at a density of $10^{6}$ cm$^{-3}$.
These conditions should be applicable to the cold, dense
cloud core of L1544.  

\subsection{Grid of Models}
A guided search of parameter space yielded a preliminary best-fit (base) 
model as judged by the $\chi^{2}$ deviation between the model and
observations (see Section 4).  
In order to confirm the best fit, a relatively fine grid 
of over 400 models were constructed surrounding this best fit model to test
the quality of fit and the geometry of the parameter space.
From the base model, sub-grids were constructed by varying pairs
of parameters:  amount 
and distribution of dust ($\tau$, $m$), the strength of heat 
sources ($G_\mathrm{MMP}$, $L_{*}$), and the source geometry
($D$, $q$).  A best fit model from these
sub-grids was identified as the new base model.  The process was
iterated multiple times to ensure the best fit was found.
A final grid was then made surrounding the best fit model, with 
the comparison to observations discussed in Section 4.
 
\section{Comparison with observations}
In order to compare the model predictions with observations, the
simulated and observed data were fit using the reduced 
$\chi^{2}$ statistic.  For the $450$ $\mu$m and $850$ $\mu$m
sky maps, the major and minor axes of the isophote contours
were compared, such that
\begin{equation}
\chi_{450}^{2} = \frac{1}{3} \sum_{i=1}^{2} \sum_{j=1}^{2} 
\left[
\frac{(r_{450,i,j,\mathrm{obs}}-r_{450,i,j,\mathrm{model}})^2}
      {\sigma_{450,i,j}^2}
\right],
\end{equation}
and 
\begin{equation}
\chi_{850}^{2} = \frac{1}{11} \sum_{i=1}^{6} \sum_{j=1}^{2} 
\left[
\frac{(r_{850,i,j,\mathrm{obs}}-r_{850,i,j,\mathrm{model}})^2}
      {\sigma_{850,i,j}^2}
\right].
\end{equation}
Here $i$ corresponds to the number of the contour level, $j$ corresponds
to the axis (major,minor), $r$ is the radius of the $i^{\mathrm{th}}$
contour along the $j^{\mathrm{th}}$ axis, $\sigma$ is the uncertainty
in the radius, 
and the leading fraction
denotes the $\frac{1}{N-1}$ term with $N_{450}=4$ and 
$N_{850}=12$ surface brightness contour data points 
(major and minor axes for each contour).
The quality of the fit of the SED is measured
similarly, via
\begin{equation} 
\chi_{\mathrm{SED}}^{2} = \frac{1}{5} \sum_{k=1}^{6}
\frac{(F_{k,\mathrm{obs}}-F_{k,\mathrm{model}})^2}{\sigma_{k}^2}.
\end{equation}
Here $k$ specifies the flux data point, $F_k$ is the flux of the
$k^{\mathrm{th}}$ datapoint, $\sigma_k$ is the uncertainty
in that flux, 
and the leading
$\frac{1}{5}$ denotes the $\frac{1}{N-1}$ term with 
$N_{\mathrm{SED}}=6$ flux data points.  
The observed data are taken from Table 1.

The overall quality of fit was then taken from 
the average of the three reduced chi-squares (assuming that
the 450 $\mu$m, 850 $\mu$m, and SED results are independent) as
\begin{equation}
\chi_{\mathrm{tot}}^{2} = 
(\chi_{450}^{2} + \chi_{850}^{2} + \chi_{\mathrm{SED}}^2) / 3.
\end{equation}
This choice for overall fit quality is advantaged by the fact
that in general none of the individual $\chi^{2}$'s dominate
the others for good fits.  
Furthermore, the individual $\chi^{2}$'s are potentially more
sensitive to different parameters.  For example, the SED is 
dependent on the dust mass (e.g., Hildebrand 1983; Doty \& Leung 1994),
while the surface brightness is more dependent upon the density 
and temperature distribution (e.g. Shirley, Evans, \& Rawlings
2002; Doty \& Palotti 2002).
As a result, the fit to $\chi_{\mathrm{tot}}^2$ 
requires a simultaneously strong fit to all data.
Unless otherwise noted, it is this value of chi-squared that is
discussed below.   For reference, our best fit model 
has $\chi^2_{\mathrm{tot}} = 0.5$.

\subsection{Optical depth and density distribution}

%
   \begin{figure}
      \resizebox{\hsize}{!}{\includegraphics{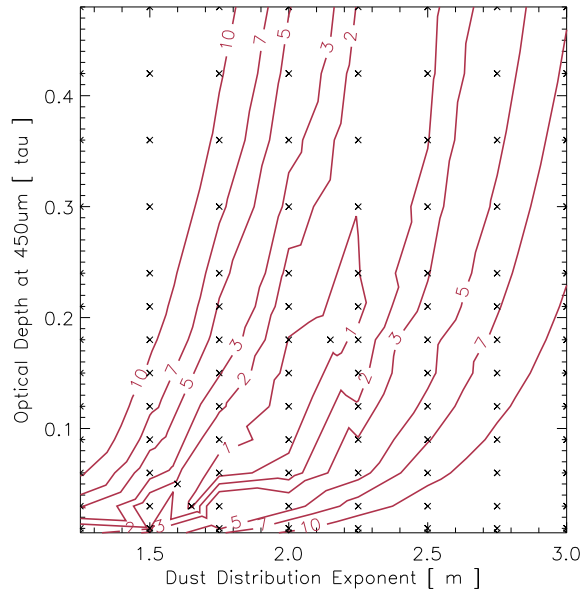}}
      \caption[]{Contours of $\chi^2$ fit for the models in the
      $\tau_{450}-m$ plane.  Note that while there is some
      degeneracy, the region is limited, enabling a constraint
      on both $\tau_{450}$, and $m$.  
              }
         \label{mtchi}
   \end{figure}

The amount of dust and its distribution can be parameterized by 
$\tau$ and $m$.  In Fig. 1, we plot $\chi^2$ contours for a grid of models in 
the $\tau_{450} - m$ plane.  The models run are denoted by the
crosses.  There exists a slight degeneracy between $\tau_{450}$
and $m$, in that a range of models yield $\chi^2 < 1$.  This
is similar to the degeneracy in the $\tau-m$ parameter space
found by Doty \& Palotti (2002) when fitting the SEDs of 
spherical models.  However, in this case, the spatial distribution
of intensity (i.e. the sky maps) allow the degeneracy to be
more nearly broken. 

Based upon these results, we infer that 
the optical depth at $450$ $\mu$m can be constrained to 
$ 0.06 < \tau_{450,\mathrm{long}} < 0.24$.  These values are, however, 
defined along the long-axis of the ellipsoid.
The optical depth along the short axis is a factor of $\sim 4$ lower, or in
the range $ 0.015 < \tau_{450,\mathrm{short}} < 0.06$.  This corresponds
to optical depths of 
$2 \times 10^{-3} < \tau_{1300,\mathrm{short}} < 8 \times 10^{-3}$ at 
$1300 \mu$m, and 
$0.06 < \tau_{200,\mathrm{short}} < 0.24$ at $200 \mu$m.
These ranges are consistent with those of previous observations
and spherical modeling.
This is to be somewhat expected, as at these wavelengths the 
source is transparent, so that nearly
all grains can be seen.  As a result, the total amount
of energy absorbed from the ISRF plays a slightly larger
role than the spatial symmetry, meaning that the spherical
model `average optical depth' is an appropriate starting point.

The results of Fig. 1 also suggest that the dust density
distribution can be well fit by a power law of the form 
in equation (1), with an exponent $1.75 < m < 2.25$.  This
is generally consistent with the previous results from
spherical modeling which suggest that a 
power law having $m=2$ would fit the data
(Evans et al. 2001), so long as a self-consistent
temperature distribution is adopted.  Again, this 
is somewhat expected, as the density powerlaw can be viewed as
essentially fixing the relative amounts of `warm' outer dust to 
`cold' inner dust.  Consequently, for a relatively regular
source which is transparent at the wavelengths of interest, it 
appears that the spherical average density distribution provide a
reasonable first-order estimate.  Most strikingly,
when a self-consistent, depth-dependent temperature distribution, and
the source asphericity are taken into account, the data are well-fit
by a singular power-law density distribution.  

In order to probe the ability of a Bonner-Ebert-like density
power law to fit the observations, we have also considered
models with a flattened density distribution within a sphere of radius
$r_\mathrm{cut}$.   When we allow the density distribution for
$r>r_{\mathrm{cut}}$ to vary (keeping $m$, and $\tau_{450}$ constant), 
we find that a central
flattening for $r<0.002$ pc $=400$ AU will produce adequate fits.
On the other hand, when the density profile is unchanged 
for $r>r_{\mathrm{cut}}$ (allowing $\tau_{450}$ to vary), acceptable fits
are found for $r_{\mathrm{cut}} < 0.008$ pc $= 1600$ AU.  
This is roughly consistent with the results of
Evans et al. (2001) in the sense that 
our density at $1600$ AU is $\sim 10^{6}$ cm$^{-3}$, the same as the
central density for their best-fitting BE sphere.
In order to place $r_{\mathrm{cut}}<1600$ AU in context, we note that the
size of the $\sim 15$ arcsec beam at $850\mu$m is $\sim 0.01$ pc = $2000$ AU.
As a result, while we cannot rule out a flattened Bonner-Ebert-like
central density distribution, it is restricted to $r<1600$ AU, and
is essentially unresolved / at the resolution limit of the observations.

%
   \begin{figure}
      \resizebox{\hsize}{!}{\includegraphics{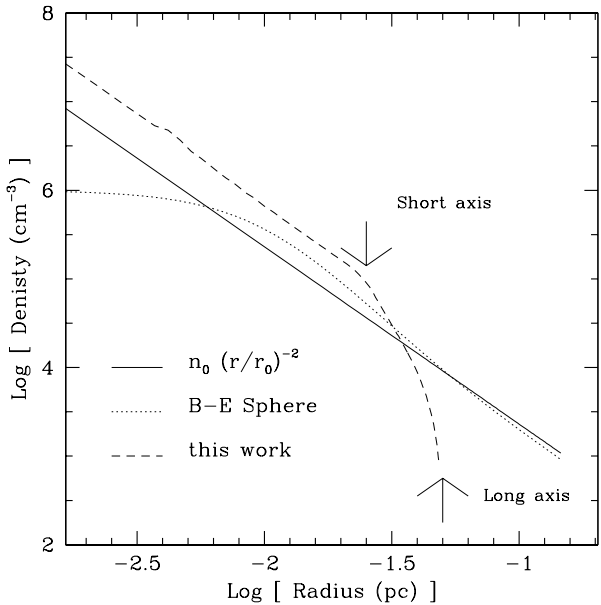}}
      \caption[]{Density profiles for L1544.  The spherical
      average of the best-fit ellipsoidal distribution
      is given by the long-dashed lines.  Also plotted
      are the corresponding
      best-fit Bonner-Ebert sphere (dotted line) and 
      spherical power law (solid line) from Evans et al. (2001).  
              }
         \label{densplot}
   \end{figure}

The best-fit density distribution and spherical comparators
are given in Fig. 2.  The solid and dotted lines are the best-fit
power law and Bonner-Ebert spherical density profiles from
Evans et al. (2001).  The long-dashed line is the spherical average
of our best-fit ellipsoidal profile (see Sect. 3.1).  Notice that,
in general, our best-fit model represents a denser, smaller cloud
than the previous spherical modeling.  
Furthermore, our best-fit model maintains an $m=2$ profile, for 
$r < R_{\mathrm{short}}$.  However, once $r>R_{\mathrm{short}}$, 
the spherical average density falls much faster with $m \sim 4-5$ due to the
averaging of high density on-cloud and low-density off-cloud
positions in taking the spherical average.  Interestingly,
this steep density fall-off 
is consistent with the results of Abergel et al. (1996)
and Bacmann et al. (2000), even though along any given ray
from the center of the cloud, $m=2$.

Finally, we note that our temperature distribution is roughly 
consistent with the self-consistent spherical modeling of 
Evans et al. (2001), and the analytic work of Zucconi, 
Walmsley, \& Galli (2001) for L1544.  In particular, we find
an unattenuated (outer) spherical average dust temperature of $\sim 15$ K, 
in keeping with the work of Zucconi et al. (2001), and an 
inner spherical average temperature of $\sim 5.4$ K.  The
inner temperature is consistent with the Zucconi results for
a high central density ($n_{\mathrm{H}} = 4.4 \times 10^{6}$ cm$^{-3}$, 
but approximately 2 K lower than the Evans et al. (2001) and
Zucconi et al. (2001) results for power-law density distribution
adopted here.  This has two causes.  First, the central density 
for our power law model is closer to that of Zucconi et al. (2001).  
Also, while we our short axis is only slightly more opaque than the
spherical models of Evans et al. (2001),  
our long axis is more opaque by a factor of $\sim 2$.
The increased opacity leads to less radiation pentration, and
thus lower temperatures in our models.

\subsection{ISRF and central luminosity}

%
   \begin{figure}
      \resizebox{\hsize}{!}{\includegraphics{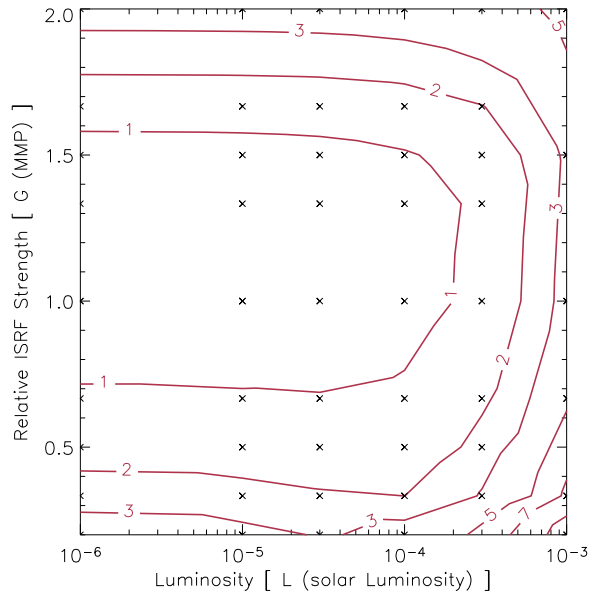}}
      \caption[]{Contours of fit $\chi^2$ in the
      ISRF-$L_*$ plane.  Note that there is no degeneracy
      between the parameters, and that significant limits can
      be set.
              }
         \label{ilchi}
   \end{figure}

The strength of the external and internal heat sources are 
parameterized by $G_{\mathrm{MMP}}$ and $L_{*}$.
In Fig. 3, we plot the $\chi^2$ contours in the 
$G_{\mathrm{MMP}}$-$L_*$ plane.  
Again, the crosses signify the models considered.
In this case, we can see that there is no degenercy between 
these two parameters (similar to results for fitting 
SEDs, Lis et al. 2001).  This is due to the fact that the 
size and spacing of the outer contours is mostly determined
by the external heat source (the ISRF), while the 
inner contours are determined by the opacity at the wavelength 
of the dominant heating radiation and the luminosity of the
central source.  Consequently, each parameter probes a different
region of the spatial intensity distribution maps.

Based upon the results of Fig. 3, we can constrain the strength of the 
external heat source, namely the ISRF, 
to be $0.8 < G_{\mathrm{MMP}} < 1.6$.   This is significant
in that while it is consistent with the MMP 
ISRF, it is approximately a factor of 1.5-3 larger than the reduced 
field used by Evans et al. (2001),  
and the attenuated field adopted by Young et al. (2004)
in their spherical modeling.   This difference is due to the fact that for
wavelengths which dominate the absorbed radiation ($\lambda < 25 \mu$m),
the spherical clouds are effectively larger (more optical depth at
larger radii).  Thus, the stronger radiation field found here   
is consistent with the smaller surface area, mass, and
higher optical depth in our model versus the equivalent 
spherical modeling.  

Last, it is useful to note that 
we have considered the $G_{\mathrm{MMP}}-D$ plane
of fits, finding that $G_{\mathrm{MMP}}$ is relatively unaffected due to the 
small optical depth of the models.  Together with the discussion above, this
result points to the potential to infer local conditions when modeling 
combined spectral and spatial data.

Of perhaps even greater interest, it is also possible to constrain
the central luminosity.  We find  $L_{*} < 10^{-3}$ L$_{\odot}$.
This result is commensurate with the fact that no IRAS source exists
at this position (Beichman et al. 1986), and with the assignation of
this source as a preprotostellar object (Ward-Thompson,
Scott, Hills, \& Andr\'{e} 1994).  The dominant factor in 
constraining the luminosity is the surface brightness distribution.
For the source mass and optical depth, a central source of
luminosity $L > 10^{-3} L_{\odot}$ yields a relatively strong
point source in the maps, and a concomitant decrease in the
size and ellipticity of the intensity contours.

The luminosity constraint at first appears inconsistent with the central
luminosity quoted in Table 1.  However, the limit in Table 1 arises from 
the difference between the incident luminosity
from the ISRF ($\sim 0.1 L_{\odot}$) and the emergent luminosity 
($\sim 0.2 L_{\odot}$) found by Ward-Thompson, Andr\'{e}, \& Kirk (2002).  
As discussed by those authors, the small difference suggests that 
$L_{\mathrm{central}} \sim 0$ may well be within the
uncertainties in their analysis.   Consequently, the work here 
provides a much more strenuous constraint on the upper limit to the
luminosity of any central source. 
Furthermore, while there may be
infall (Tafalla et al. 1998; Williams et al. 1999), 
the low luminosity inferred here 
confirms the previous result that the collapse is nearly isothermal
with a correspondingly small compressional luminosity (Henriksen 1994), 
and is inconsistent with the formation of a luminous protostar
at the center of the collapsing core.

\subsection{Outer diameter and axis ratio}

%
   \begin{figure}
      \resizebox{\hsize}{!}{\includegraphics{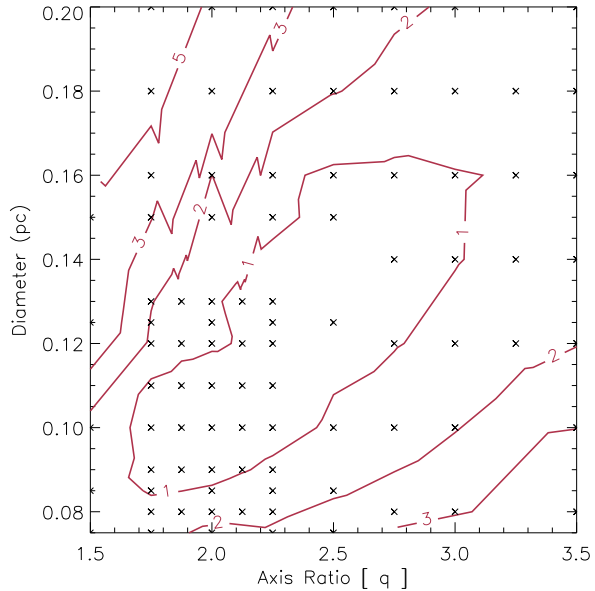}}
      \caption[]{Contours of $\chi^2$ fit for models in the
      $D - q$ plane.  Note that while there is
      a degeneracy, it is limited so that constraints can
      be placed on these parameters.
              }
         \label{qrchi}
   \end{figure}

The size and shape of the cloud are parameterized by $D$, and $q$.
The $\chi^2$ contours for the diameter ($D$) - 
axis ratio ($q$) plane are shown in Fig. 4.  While there is some
degeneracy between the diameter and axis ratio, it is 
again limited.  In this case, the degeneracy is limited by the
fact that the flux and hence the SED is more sensitive to the
outer diameter, while the spatial intensity contours are sensitive
to both the axis ratio and the diameter.
 
Based upon the results in Fig. 4, 
the outer diameter of the long axis of the
source can be constrained to $0.08 < D(\mathrm{pc}) < 0.16$
($16000 < D(\mathrm{AU}) < 32000$).
This is consistent with the size inferred by Bacmann et al.
(2000).  It is however, smaller than the diameter adopted in the
spherical modeling of Evans et al. (2001), who adopted an
arbitrarily large diameter to simulate chopping into an
extended cloud. 
It should be noted that the actual cloud size may be somewhat
larger than the value inferred here as discussed in Section 3.

It is also possible to constrain the axis ratio of the source.
We find $1.7 < q < 2.5$.  This is roughly consistent
with the isophotes for the 450 $\mu$m and 850 $\mu$m sky
maps, which show axis ratios in the range of $1.5 - 4$.
The observed axis ratios are large because the external heating radiation
must travel a longer physical distance along the short axis
than along the long axis to reach the same optical depth.  As a
result, isophotes are more elongated than the underlying 
isodensity contours.

\subsection{Comparison of best-fit model with observations}

%
   \begin{figure}
       \includegraphics[height=84.7mm,width=84mm]{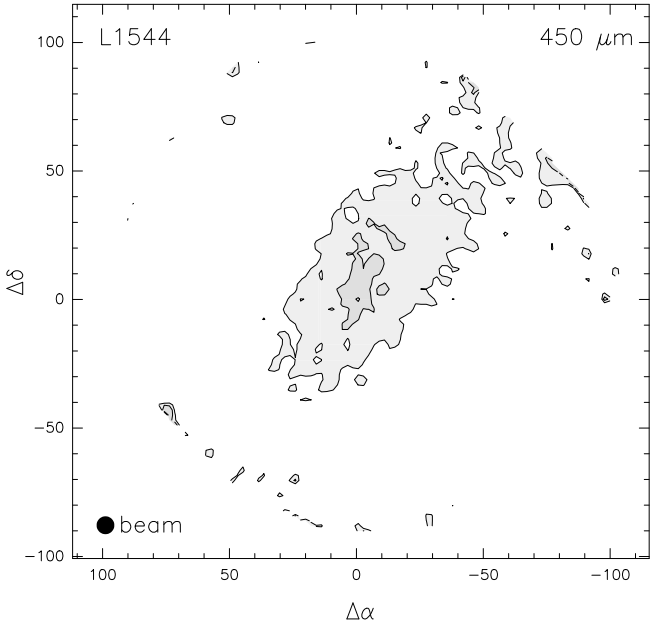}
       \includegraphics[height=85mm,width=84mm]{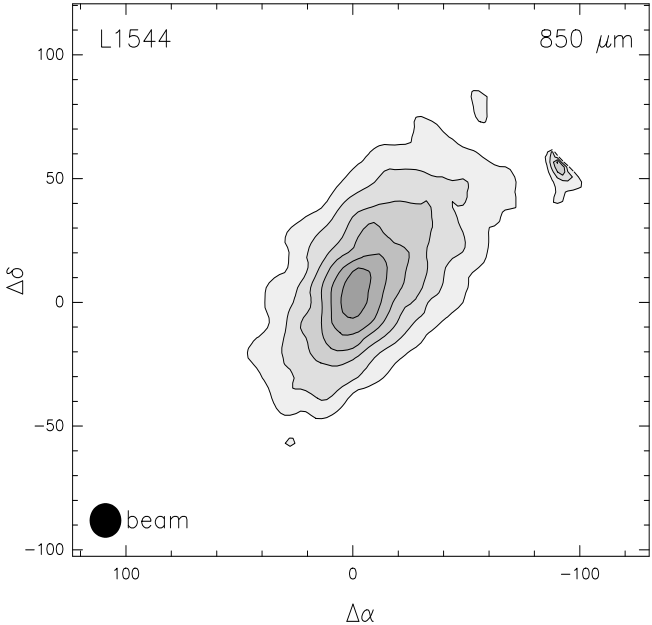}
      \caption[]{Intensity contours for L1544.
      Top panel:  450 $\mu$m; lowest contour level at
      50\% of peak flux (3 $\sigma$), with levels increasing by 33\%.
      Lower panel:  850 $\mu$m; lowest contour level at 20\% of
      peak flux (3 $\sigma$), with levels increasing by 13\%.
      Data from Shirley, Evans, Rawlings, \& Gregersen (2000).
              }
         \label{450comparison}
   \end{figure}

%
   \begin{figure}
       \includegraphics[height=84mm,width=84mm]{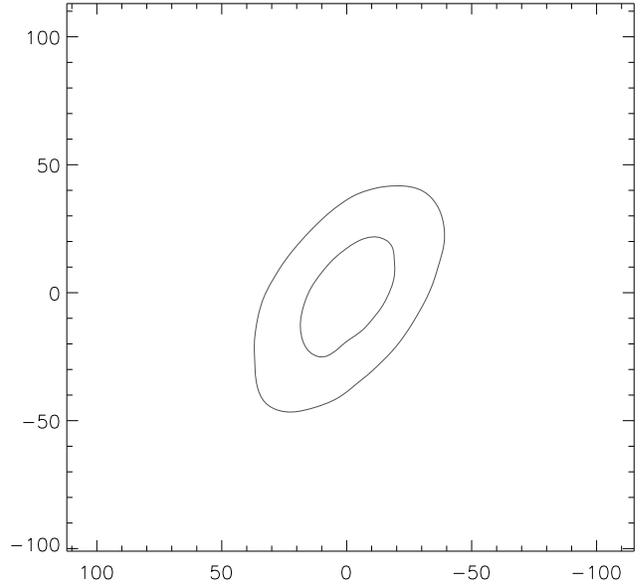}
       \includegraphics[height=86mm,width=84mm]{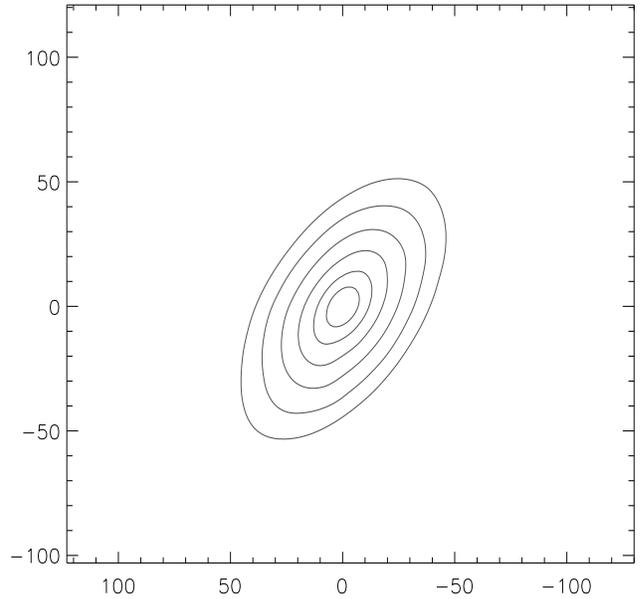}
      \caption[]{Intensity contours for L1544 at for the best fit
      model.  
      Top panel: 450 $\mu$m.  Lower panel:  850 $\mu$m.
      The contour levels are the same as in Fig. 5.
              }
         \label{850comparison}
   \end{figure}

Based upon the results above, our best fit model
for L1544 is a prolate spheroid having an axis ratio of $q=2$,
with an ellipsoidal power-law density distribution 
given by equation (1) having
$m=2$.   We infer the optical depth along the long-axis at 
$450$ $\mu$m to be $\tau_{450}=0.12$, and the central luminosity
to be $L_* = 0$.  Finally, we infer that an unscaled
MMP ISRF, and an outer diameter along the long axis of
$D=0.1$ pc = $20000$ AU (though it could be somewhat larger) provide a best
fit to the data.  

%
   \begin{figure}
     \includegraphics[height=84mm,width=84mm]{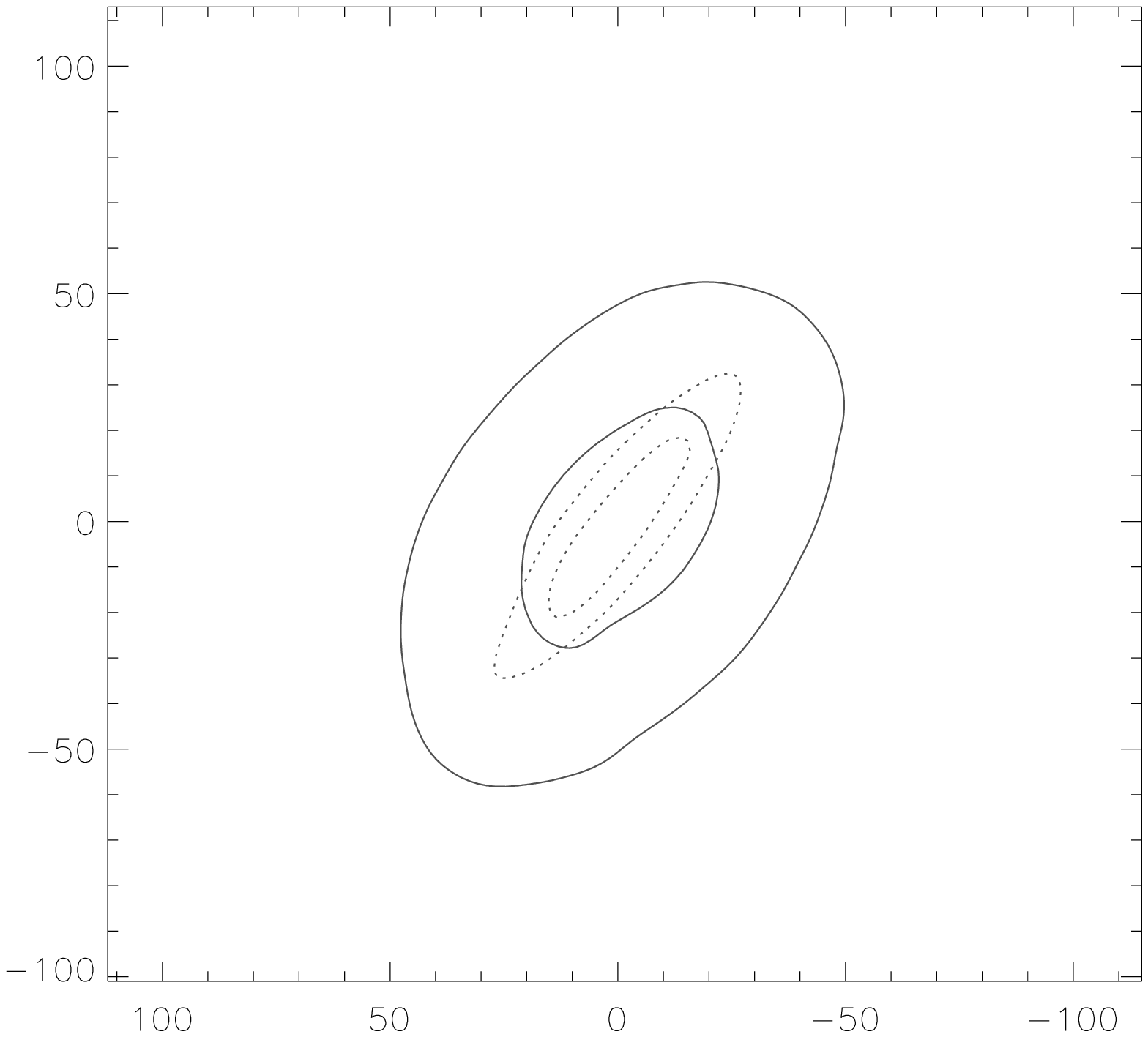}
     \includegraphics[height=86mm,width=84mm]{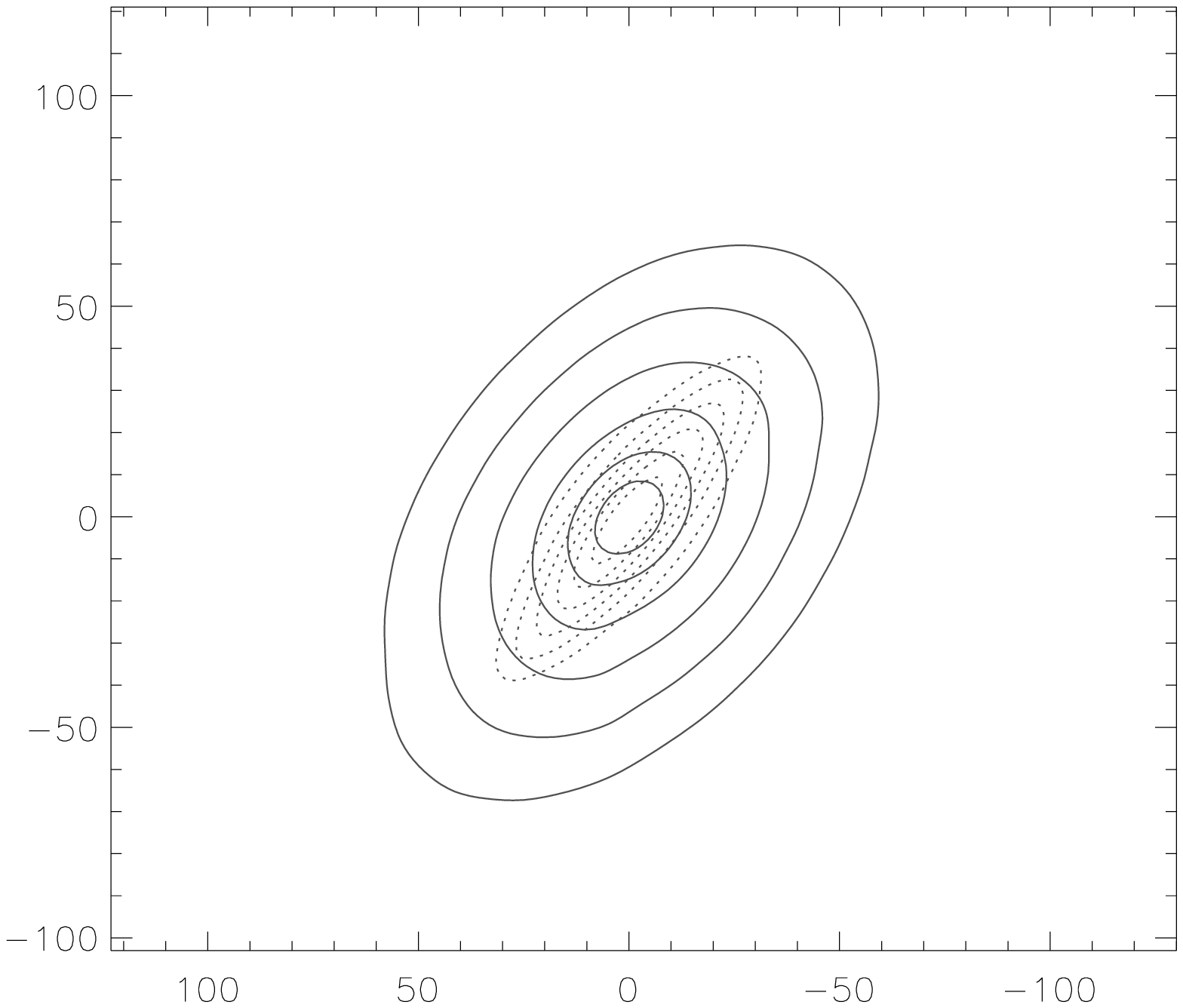}
      \caption[]{Model intensity contours for L1544 at 450$\mu$m
      (top panel) and 850$\mu$m (bottom panel)
      for a representative model having $\chi^{2}_{\mathrm{tot}}=2$
      (solid lines) and $\chi^{2}_{\mathrm{tot}}=5$ (dotted lines).
      The contour levels are the same as those in Fig. 5.
              }
         \label{chisq2sky450}
   \end{figure}

A comparison of the $450$ and $850$ $\mu$m sky maps with 
the best fit model is shown in Fig. 6.  
Notice that the fits are generally quite good, in terms of the
size and width of the intensity contours.  
For comparison, Fig. 7 gives similar sky-map intensity contours
for representative models with $\chi^{2}_{\mathrm{tot}}=2$ (solid lines), 
and $5$ (broken lines) respectively.  The contour levels are the
same as in Fig. 5.  Notice that the sky maps for even these
somewhat low values of $\chi^{2}_{\mathrm{tot}}$ are significantly 
worse than for the best fit model(s).  

Likewise, the observed and modeled SEDs are compared in Fig. 8.  
The observational data (symbols with error bars) are taken
from Table 1.  The solid line corresponds to the best-fit model, 
while the dotted and dashed lines correspond to the
$\chi^{2}_{\mathrm{tot}}=2,5$ models discussed above respectively.
As before, the best fit model reproduces the observed data well.
On the other hand, although the $\chi^{2}_{\mathrm{tot}}=5$ model 
is well-outside the observed SED limits, the same is not true for
the $\chi^{2}_{\mathrm{tot}}=2$ model.  While the 
$\chi^{2}_{\mathrm{tot}}=2$ model is at the very limits of
acceptability, it is still consistent with the observed SED.  This
is in keeping with the result of Doty \& Palotti (2002) that fitting
the SED alone can yield degenerate solutions
(see also, e.g., Men'shcikov \& Henning 1997; Lis et al. 2001).  
Furthermore, it underlines the significance of resolving the 
source, and simultaneously fitting both the surface 
brightness and SED to best constrain the source properties.

%
   \begin{figure}
      \resizebox{\hsize}{!}{\includegraphics{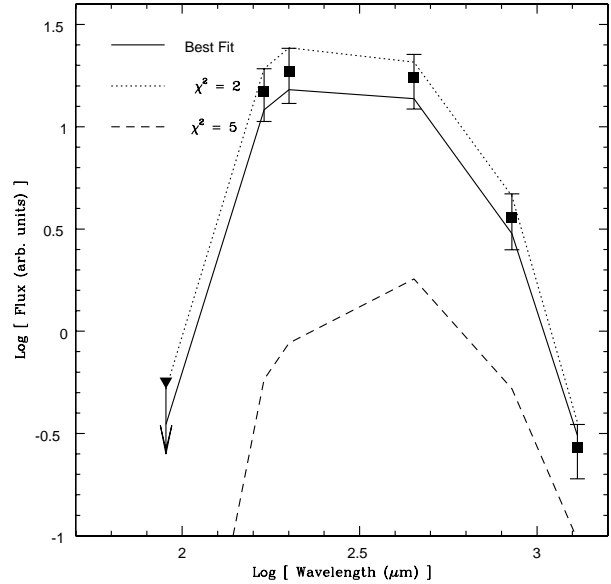}}
      \caption[]{Comparison of the SED for the best-fit
      model (solid line), a model with
      $\chi^{2}_{\mathrm{tot}}=2$ (dotted line), 
      $\chi^{2}_{\mathrm{tot}}=5$ (dashed line),
      and observations (solid symbols with errorbars).
              }
         \label{SEDcomparison}
   \end{figure}

Finally, the inferred ranges and best fit values for the
parameters specifying the structure and conditions of L1544
are summarized in Table 3, along with values inferred from previous
work.  As noted previously, we see that conclusions from
earlier modeling of observations at wavelengths for which the
source is optically thin yield results which are relatively close
to the best fit models.  Even more interesting is the fact that
under the same conditions, spherical models yield results which 
are reasonable first-order approximations.  Still, it should be
noted that even in the simple case of a centrally condensed, 
externally heated, optically thin, ellipsoidal power law density
distribution with an axis ratio of $\sim 2$, geometrical and
radiative transfer effects do cause deviations from results 
inferred from simpler analyses.    As a result, self-consistent,
multi-dimensional modeling is an important tool for understanding 
radiation from aspherical sources.

   \begin{table}
      \caption[]{Inferred conditions for L1544}
         \label{l1544aconditions}
         \begin{tabular}{lrr}
            \hline
	        {}    & Previous & This \\
            Parameter & Value    & Work \\ 
            \hline
	     Optical depth ($\tau_{1300}$) [$\times 10^{-3}$] & $ 2 $ & $ 5 \pm 3  $ \\
	     Optical depth ($\tau_{200}$) & $ 0.06 $ & $0.15 \pm 0.09$   \\
	     Density power law ($m$) & $2^{a}$ & $2.0 \pm 0.25$  \\
	     ISRF ($G_{\mathrm{MMP}}$) & 0.6 & $ 1.0 \pm 0.5$  \\
             Central luminosity ($L_*$) [$L_{\odot}$] & $ < 0.1 $ & $< 10^{-3}$ \\
             Diameter ($D$) [$\times 1000$ AU] & $\sim 17.8$ & $12^b - 24^c$ \\	     
             Central mass ($M_{120}^d$) [$M_{\odot}$] & $0.4^{+0.8}_{-0.3}$ & $0.2 \pm 0.1$ \\
	     Total mass ($M_{\mathrm{tot}}$) [$M_{\odot}$] & 2.7 $\pm$ 0.7 & $1.4 \pm 0.9$ \\ 
             H$_{2}$ column ($N_{\mathrm{H}_{2}}$) [$10^{22}$ cm$^{-2}$] & $ 6-13$  & $14 \pm 9 $ \\
            \hline
     \end{tabular}\\
\begin{list}{}{}{}
\item[$^{\mathrm{a}}$] Bonner-Ebert sphere or Power-law with $m=2$. 
\item[$^{\mathrm{b}}$] Diameter of short axis  
\item[$^{\mathrm{c}}$] Diameter of long axis  
\item[$^{\mathrm{d}}$] In a 120 arcsec diameter beam  
\end{list}
   \end{table}

\section{Conclusions}
We have constructed models for the non-spherical protostellar core
L1544 utilizing a fully three-dimensional continuum radiative
transfer model.  After convolving the model output with the 
actual SCUBA beam profile, we are able to compare our results with
existing observations, including both SEDs and 
sky maps.  Based upon this work, we find that:

1.  It is possible to constrain the optical depth and density 
distribution.  While a degeneracy in this plane exists as noted
by Doty \& Palotti (2002), the spatial intensity distribution 
more nearly breaks the degeneracy.  In particular, we find 
$2 \times 10^{-3} < \tau_{1300,\mathrm{short}} < 8 \times 10^{-3}$, and
$0.06 < \tau_{200,\mathrm{short}} < 0.24$.  Likewise, we find that
an ellipsoidal density power law with an exponent $m \sim 2$ 
$(1.75 < m < 2.25$), can fit the observations well.
While a Bonner-Ebert-like centrally flattened density distribution
cannot be ruled out, the central flattening can be restricted to
$r<0.008$ pc $=1600$ AU (Sect. 4.1).

2.  The ISRF and central luminosity are nearly orthogonal parameters.
The ISRF can be constrained to be within $\sim 50$ per cent of that
defined by Mathis, Mezger, \& Panagia (1983), somewhat higher than previously
inferred from spherical modeling.  The central luminosity
can also be constrained to $< 10^{-3}$ L$_{\odot}$, consistent 
with the assignation of this source as a pre-protostellar object
(Sect. 4.2).  

3.  The diameter and axis-ratio show a limited degeneracy, which is 
again broken by the spatial intensity distribution.  We infer 
a long-axis diameter of $0.08 < D(\mathrm{pc}) < 0.16$
($16000 < D(\mathrm{AU}) < 32000$), and an axis
ratio $1.7 < q < 2.5$.  
The outer diameter and axis ratio may each be somewhat larger due
to potential on-source chopping in the observations, 
and the projection of the long axis onto the plane of the sky.  
Both the diameter ($D$) and the axis ratio ($q$) 
are less than would be inferred directly
from observations, and are due to radiative transfer effects and
non-spherical geometry (Sect. 4.3).

4.  The values and ranges of values found here
are roughly consistent with those inferred directly from observations
or spherical modeling, due to the relative transparency of the source
at the wavelengths of observation.  In general, we infer a 
smaller source of lower mass and higher optical depth / column density
than previously assumed (Sect. 4.4).  

5.  The SED yields degenerate fits.  A ``best-fit'' requires both
the SED and surface brightness distribution.  This approach, combined
with a large grid of models ($\sim 400$), can yield a best-fit and
range of uncertainty in inferred parameters.  Interestingly, a range
of 10 ($0.5 - 5$) or even 4 ($0.5 - 2$) in $\chi^{2}$ can yield 
remarkably different quality of fit (Sect. 4.4).

\section*{Acknowledgements}
      We are grateful to Chad Young and Rebecca Metzler
      for interesting discussions.
      This work was partially supported under
      grants from The Research Corporation and the
      Denison University Research Foundation (SDD), 
      NASA grants NAG5-10488 and NNG04GG24G to the
      University of Texas at Austin (NJE), and
      a Battelle Intership (MLP).

\label{lastpage}


\begin{thebibliography}{}

   \bibitem[1996]{Abergeletal1996} Abergel, A. et al. 1996, A\&A, 315, L329

   \bibitem[1991]{Adams1991} Adams, F. C. 1991, ApJ, 382, 544

   \bibitem[1996]{AndreWardThompsonMotte1996} Andr\'{e}, P., 
      Ward-Thompson, D., \& Motte, F. 1996, A\&A, 314, 625

   \bibitem[2000]{Bacmannetal2000} Bacmann, A., et al. 2000,
      A\&A, 361, 555
      
   \bibitem[1986]{Beichmanetal1986} Beichman, C. A.,  
      et al. 1986, ApJ, 307, 337           

   \bibitem[1989]{BensonMyers1989} Benson, P. J., \& Myers, P. C.
      1989, ApJS, 71, 89

   \bibitem[1991]{butneretal1991} Butner, H. M., Evans II, N. J.,
      Lester, D. F., Levreault, R. M., \& Strom, S. E. 1991,
      ApJ, 376, 636 
      
   \bibitem[2000]{CiolekBasu2000} Ciolek, G. E., \& Basu, 
      S. 2000, ApJ, 529, 925      
      
   \bibitem[2001]{CurryStahler2001} Curry, C. L., \& 
      Stahler, S. W. 2001, ApJ, 555, 160

   \bibitem[1994]{DotyLeung1994} Doty, S. D., \& Leung, C. M.
      1994, ApJ, 424, 729
      
   \bibitem[2002]{DotyPalotti2002} Doty, S. D., \& Palotti, M. L. 2002, 
      MNRAS, 335, 993
            
   \bibitem[1988]{EganLeungSpagna1988} Egan, M. P., Leung, C. M.,
      \& Spagna, G. F., Jr. 1988, Comput. Phys. Comm., 
      48, 271

   \bibitem[1978]{Elias1978} Elias, J. H. 1978, ApJ, 224, 857
      
   \bibitem[2001]{EvansRawlingsShirleyMundy2001} Evans II, N. J.,
      Rawlings, J. M. C., Shirley, Y. L., \& Mundy, L. G. 2001, 
      ApJ, 557, 193
      
   \bibitem[1994]{Henriksen1994} Henriksen, R. 1994, in Montmerle
      T., Lada C. J., Mirabel I. F., Tran Than Van J., eds., The
      Cold Universe, Editions Frontieres, Gif-sur-Yvette, p. 241    

   \bibitem[1983]{Hildebrand1983} Hildebrand, R. H. 1983, QJRAS, 24, 267
      
   \bibitem[1991]{Laddetal1991} Ladd, E. F., Adams, F. C., 
      Fuller, G. A., et al. 1991, ApJ, 382, 555L      
      
   \bibitem[2001]{Lisetal2001} Lis, D. C., Serabyn, E., Zylka, R.,
      \& Li, Y. 2001, ApJ, 550, 761      

   \bibitem[1983]{MMP} Mathis, J. S., Mezger, P. G., \&
      Panagia, N. 1983, A\&A, 128, 212
 
   \bibitem[1997]{MenshcikovHenning1997} Men'shcikov, A. B., \& Henning,
      Th. 1997, A\&A, 318, 879
      
   \bibitem[2002]{Muelleretal2002} Mueller, K. E., Shirley, Y. L., 
      Evans, N. J. II, \& Jacobson, H. R. 2002, ApJS, 143, 469    
      
   \bibitem[1983]{MyersLinkeBenson1983} Myers, P. C., Linke, R. A., 
      \& Benson, P. J. 1983, ApJ, 264, 517      

   \bibitem[1991]{Myersetal1991} Myers, P. C., Fuller, G. A., 
      Goodman, A. A., \& Benson, P. J. 1991, ApJ, 376, 561

   \bibitem[2002]{NakamuraLi2002} Nakamura, F., Li, \& Z.-Y. 2002, 
      ApJ, 566, 101L
      
   \bibitem[1994]{OssenkopfHenning1994} Ossenkopf, V., \&
      Henning, Th. 1994, A\&A, 291, 943   
            
   \bibitem[1996]{Ryden1996} Ryden, B. S. 1996, ApJ, 471, 822      

   \bibitem[2002]{ShirleyEvansRawlings2002} Shirley, Y. L., 
      Evans II, N. J., \& Rawlings, J. M. C. 2002, ApJ, 575, 337

   \bibitem[2000]{Shirleyetal2000} Shirley, Y. L., Evans II, N. J., 
      Rawlings, J. C., \& Gregersen, E. M. 2000, ApJS, 131, 249   

   \bibitem[1991]{SpagnaleungEgan1991} Spagna, G. F. Jr., Leung, C. M.,
      \& Egan, M. P. 1991, ApJ, 379, 232
      
   \bibitem[1998]{Tafallaetal1998} Tafalla, M., Mardones, D., 
      Myers, P. C., et al. 1998, ApJ, 504, 900    
      
   \bibitem[1999]{vdt99} van der Tak, F. F. S., et al. 1999,
      ApJ, 522, 991
      
   \bibitem[2000]{vdt2000} van der Tak, F. F. S., et al. 2000, 
      ApJ, 537, 283 
      
   \bibitem[2002]{WardThompsonAndreKirk2002} Ward-Thompson, D., 
      Andr\'{e}, P., \& Kirk, J. M. 2002, MNRAS, 329, 257      
      
   \bibitem[1999]{WardThompsonMotteAndre1999} Ward-Thompson, D.,
      Motte, F., \& Andr\'{e}, P. 1999, MNRAS, 305, 143           

   \bibitem[1994]{WardThompsonScottHillsAndre1994} Ward-Thompson, D.,
      Scott, P. F., Hills, R. E., \& Andr\'{e}, P. 1994, MNRAS, 268, 276   

   \bibitem[1999]{Williamsetal1999} Williams, J. P., Myers, P. C., 
      Wilner, D. J., \& di Francesco, J. 1999, ApJ, 513, L61

   \bibitem[2003]{YoungShirleyEvansRawlings2003} Young, C. H., 
      Shirley, Y. L., Evans, N. J., \& Rawlings, J. M. C. 2003, 
      ApJS, 145, 111
      
   \bibitem[2004]{Youngetal2004} Young, K. E., Lee, J.-E., 
      Evans II, N. J., Goldsmith, P. F., \& Doty, S. D. 2004, 
      ApJ, in press        
      
   \bibitem[2001]{Zucconietal2001} Zucconi, A., Walmsley, C. M., 
      \& Galli, D. 2001, A\&A, 650, 662            


\end{thebibliography}
\end{document}